# Twitter adoption, students' perceptions, Big Five personality traits and learning outcome: Lessons learned from 3 case studies

## Alexia Katrimpouza, Nikolaos Tselios and Maria-Christina Kasimati

ICT in Education Group, Dept. of Educational Sciences and Early Childhood Education
University of Patras
Rio, Patras, 26500, Greece
alexiakatrimpouza@gmail.com , nitse@ece.upatras.gr, kasimatimx@gmail.com

**Abstract:** This study examines the relationship of the Twitter's use with the participants' learning outcome through a series of well-organized educational activities. Three studies were conducted in the context of two academic courses. In all three studies the students who participated in the process received a higher laboratory grade than the students who did not participated. Students' conscientiousness and openness to experience were related to their activity in one study. However, no relationship between the students' personality traits and their grade was unveiled. Moreover, the students' interventions in the activities are examined as well as the differentiation in their attitudes towards social media use in learning. The implications of the results are discussed and a comparison with other related studies is presented.

## Introduction

Web 2.0 applications offer significant opportunities in education due to their open nature, their inherent ease of use and the support they provide for active participation and cooperation of the users. (Altanopoulou., Tselios, Katsanos, Georgoutsou, & Panagiotaki, 2015; Crook, 2008; Tselios, Altanopoulou & Komis, 2011). Twitter is a Web 2.0 technology which allows micro-blogging: that is, it allows users to create posts with a limited number of characters. Twitter is considered as an acceptable and useful tool for direct communication among peers (Hattem, 2012, Straus, Williams, Shogan, & Glassman, 2014). Recent studies demonstrated that it also improves active learning and enhances learning incentives (Chen & Chen, 2012; Domizi, 2013; Wakefield, Warren, & Alsobrook, 2011). Students gain confidence, fluency in expressing ideas and they also enhance their social skills (Chen & Chen, 2012; Smith & Tirumala, 2012). By using Twitter, participation, engagement, reflection and collaborative learning is encouraged (Gao, Luo & Zhang, 2012).

In a study conducted by Tiernan (2013) with 75 first-year students, three research questions on the use of Twitter in the educational community were studied. Firstly, they examined whether the students adopted Twitter within the context of the course and on which motives. Secondly, the contribution of Twitter to the engagement of the students with the course has been studied and thirdly, the different kinds of use of the tool have been examined. Student engagement is defined as the time and effort students invest in educational activities which are linked to desired outcomes (Kuh, 2009). It was observed that students have poorly adopted the tool, reporting that there were technological barriers and lack of motivation (Tiernan, 2013). Students reported that Twitter helps their interaction in a variety of ways. Firstly, questions about the course can be submitted. Secondly, it is possible to easily share ideas. Additional advantages were the given answers related to the course and the annotations made by the students, respectively. Moreover, students reported three possible uses of the tool to the academic community: 38% of the students reported that they would be happy to use the tool outside the classroom. Subsequently, 28% of the students suggested to use Twitter as a formal newsfeed of the course. Finally, 31% of the students mentioned that they would like to use it as an additional source of communication with the teacher.

Junco, Heibergert and Loken (2010) investigated the educational use of Twitter and the influence that it appears to have on students regarding their level of engagement in the learning community. 125 students participated in total: 70 were in the experimental group and 55 were in the control group. The research shows that the educational use of Twitter helps students to increase their engagement with the course. Additionally, the encouragement to use Twitter in an educational manner also appears to have positive results to the students' scores.

Prestridge (2013) used Twitter to support freshmen and analyzed the types of interaction that arise, such as connectivity, academic culture and resourcefulness. A link between the student and the instructor was established using a conventional approach with students publishing tweets and instructors answering them. However, poor interaction among the students was observed. Moreover, academic culture was cultivated by restatement of the lectures' content, by publishing images and links and by "retweeting". During the use of this technology, students dealt with several problems regarding the adoption of academic protocols, the 140 characters limitation and the completion of multiple tasks.

Furthermore, Junco, Ilavsky and Heiberger (2013) studied the systematic introduction of Twitter in the learning process. During this study, the researchers used the National Survey of Student Engagement (NSSE), a questionnaire which assesses the perceptions of students about their involvement in the academic community. In parallel with the Twitter use, the students of the control group were using Ning, an online platform which allows users to create their own communities and social networks around specific interests with their own visual design, choice of features and member data.

Subsequently, the professors asked the students to read a book and write summaries of specific chapters. After that, compulsory and optional activities were given to them. These activities should be accomplished via Twitter and every student was requested to comment on the posts of their colleagues. Moreover, they recommended students to express questions about the course, to note important dates or to mark various events regarding the University community. During the first study, the participation through the tools was compulsory for everyone. However, during the second study, the participation was optional. The results of the first study showed that the Twitter team had higher efficiency than the control group which used the Ning tool. The results of the second study, in which the participation was optional and less organized, did not show a significant difference between the students' performance. In conclusion, Junco, Elavsky and Heiberger (2013) argue that the use of Twitter benefits the students regarding their academic engagement as well as their performance. Moreover, it seems that the professors should also be involved with the tool to influence their students' results.

However, little is known about the influence of students' non-cognitive characteristics on their learning performance in the context of technology-mediated collaborative learning. Nowadays, a strong body of research demonstrates that students' personality characteristics are significantly related to their overall academic performance (Poropat, 2009). In addition, recent studies illustrated that academic performance is linked with agreeableness, conscientiousness and openness to experience (Poropat, 2009). Moreover, O'Connor and Paunonen (2007) found that conscientiousness is strongly associated with academic performance. A study conducted with 247 British university students investigates the influence of personality characteristics on academic performance, assessed by written examinations (Chamorro-Premuzic, & Furnham, 2003). Neuroticism and extraversion was negatively related to academic achievement whereas the personality characteristic of conscientiousness was positively associated with academic performance.

The relation of students' personality characteristics with their participation in learning environments, either traditional or Web-based has been also investigated (Caspi et al., 2006). Real participants in the traditional environment found to be more extroverted, open to new experiences and more emotionally stable in comparison to the students who didn't participate in this environment. However, to the best of our knowledge, the literature does not attempt to investigate the influence of students' personality traits on Twitter adoption and the learning outcome or their attitudes toward Twitter due to those traits (Altanopoulou, & Tselios, 2015).

The above-mentioned literature review shows that research regarding the effect of Twitter use on the learning outcome is still at a primary stage. Therefore, more systematic research is needed regarding the use of the tool in a well-organized framework as well as the influence of the students' personality characteristics on Twitter's adoption and use. The present study investigates the efficiency of the activities' design framework proposed by Junco, Elavsky and Heiberger (2013) in the context of an academic course.

**Research Methodology**

*Research questions*

A one group pre-posttest design was adopted (Cohen, Manion, & Morrison, 2013). This study aims to investigate the relationship, if any, between personality traits of higher education students and the use of the social media of Twitter. It also attempts to study the higher education students' interventions in Twitter and how they are related with the learning outcome. Moreover, it seeks to design and evaluate educational activities mediated by Twitter. Finally, it attempts to examine whether there is a difference in students' attitudes towards technology before and after the implementation of the activities.

In specific, the research questions are:

1. Is there any relationship between the personality traits of higher education students and the use of Twitter?
2. Is there any relationship between the students' personality traits and their academic performance?
3. Is there any relationship between Twitter activity and students' academic performance?
4. Is there a difference between students' perceptions before and after the implementation of the educational activities via Twitter, regarding:
   - Their attitude towards learning via social media,
   - Their attitude towards learning via ICT,
   - Their self-reported attraction towards technology?

*Participants*

The participants were students of the Department of Educational Sciences and Early Childhood Education at the University of Patras, Greece. The first case study took place during the elective course "Introduction to Web Science" which was taught in the second semester (February 2015- June 2015) of the first academic year. Subsequently, two more studies were implemented. During the first semester (September 2015- January 2016) the second case study took place, within the compulsory first year course "Introduction to ICT". Finally, during the second semester (February 2016- June 2016) the third case study was implemented in the context of the elective course entitled "Introduction to Web Science".

1st study: 19 students (0 male, 19 female), participated to the research, aged 18- 28 (mean=20.03, SD=2.2). Regarding the school/academic performance of the participants: the high school graduation grade had mean = 17.4/20 (SD = 0.9), their mean laboratory grade was 9.05/10, (SD= .39), and the mean final exams grade was 7.0/10, SD= 1.19). Their mean grade in all completed courses at the time was 5.6 (SD=3.5).

2nd study: 80 students (78 male, 2 female), aged 17-47 (mean=19.0, SD=3.5). Their high school graduation grade had mean= 16.9/20 (SD= 1.5), their mean laboratory grade was 8.0/10 (SD= 1.5), and their mean final exam grade of the course was 5.1/10 (SD= 2.2).

3rd study: 46 students (1 male, 45 female), aged 18- 33 (mean=20.6, SD=2.2). Their high school graduation grade had mean = 17.4/20 (SD = 1.4), their laboratory grade was mean = 8.8/10, SD = 0.7, and the mean final exams grade was 5.6/10 (SD = 2.1). Their mean grade in all completed courses at the time was 7.9 (SD=1.2).

*Materials*

The materials used during the research process were the questionnaire and the educational activities via Twitter. The instruments used were the Social Media Learning (SML) scale, Technology Affinity Survey (TAS), Communications Technology Learning (ICTL) survey (Knezek, Mills, & Wakefield, 2012, Mills, Knezek, & Wakefield, 2013) and Big Five personality test (Goldberg, 1992). The educational activities were used throughout the research to create an interaction model between the students and the social network.

The questionnaires and the activities' documents were distributed to the students by using the SurveyMonkey service (www.surveymonkey.com) and the Google Docs service, respectively. SPSS v20 as well as Microsoft Excel were used for the data analysis. Finally, the Twittonomy tool has been used, which allows the user to review all the students' tweets related to the course's hashtag.

*Procedure and activities' description*

Regarding the involvement of students in Twitter, they had to answer in questions related to each of the laboratory class during an eight-week period, by using the course's hashtag. The students were informed at the end of every class for the procedure both verbally and by email. Subsequently, they had 3 days to answer the weekly questions. The students were asked: (a) to answer every week with two Tweets tops in every question, (b) to comment every week the answers of at least two of their fellow students, (c) to submit by the end of the semester at least two questions regarding the course. The tutors regularly posted information related to the course and provided feedback to the students as per their request.

Since the participation to the research was optional, the tutor set a motive for participation. 5 students with the highest activity in Twitter have been rewarded with up to 1.5 point in the final grade of the course.

**Results**

**RQ1: relationship between the personality traits of higher education students and the use of Twitter**

The Big Five questionnaire was used which includes the following personality traits: Extraversion, agreeableness, conscientiousness, emotional stability, openness to experience (Goldberg, 1992). The present research uses the Greek version of the instrument. The activity on Twitter was assessed by the total number of tweets post by the participants during each study.

Regarding the first and the second study, no statistically significant relationships were found (See Table 1). In the third study, conscientiousness has a statistically significant relationship with the tweets that express questions towards the tutor $p= 0.009$, s, as well as with the activities' tweets $p=0.026$, s (Table 1). Moreover, openness to experience is shown to have a relationship with the activities' tweets ($r=0.395$, $p=0.007$, s).

**Table 1**. Relation between personality traits and Twitter activity (1st study: N=19, 2nd study: N=80, 3rd study: N=46.)

|  |  | Total actions | Questions | Activities | comments |
|---|---|---|---|---|---|
| Extraversion (1st study) | r | -.198 | -.065 | .258 | -.250 |
|  | p | .416 | .792 | .287 | .302 |
| Extraversion (2nd study) | r | .117 | .052 | .022 | .109 |
|  | p | .300 | .649 | .843 | .334 |
| Extraversion (3rd study) | r | -.036 | -.155 | -.062 | -.033 |
|  | p | .813 | .303 | .684 | .827 |
| Agreeableness (1st study) | r | -.138 | -.165 | -.178 | -.116 |
|  | p | .574 | .500 | .465 | .636 |
| Agreeableness (2nd study) | r | .078 | .045 | .025 | .025 |
|  | p | .494 | .690 | .827 | .827 |
| Agreeableness (3rd study) | r | .053 | .067 | .123 | .049 |
|  | p | .727 | .657 | .417 | .745 |
| Conscientiousness (1st study) | r | .103 | .236 | .052 | .092 |
|  | p | .673 | .331 | .831 | .706 |
| Conscientiousness (2nd study) | r | .088 | -.002 | .147 | .067 |
|  | p | .439 | .986 | .192 | .554 |
| Conscientiousness (3rd study) | r | .258 | .379 | .328 | .250 |
|  | p | .083 | **.009** | **.026** | .094 |
| Emotional Stability (1st study) | r | .077 | .293 | .257 | .031 |
|  | p | .755 | .224 | .288 | .900 |
| Emotional Stability (2nd study) | r | .109 | -.013 | .202 | .084 |
|  | p | .335 | .906 | .072 | .460 |
| Emotional Stability (3rd study) | r | -.086 | .101 | .060 | -.092 |
|  | p | .569 | .505 | .690 | .544 |
| Openness to experience (1st study) | r | -.101 | .072 | .016 | -.117 |
|  | p | .682 | .770 | .947 | .634 |
| Openness to experience (2nd study) | r | .009 | .068 | .094 | -.005 |
|  | p | .934 | .547 | .409 | .964 |
| Openness to experience (3rd study) | r | .035 | .207 | .395 | .019 |
|  | p | .815 | .168 | **.007** | .901 |

**RQ2: relationship between the students' personality traits and their academic performance**

Data analysis in the first study showed that the students' agreeableness had a relationship with their final grade $r= -.40$ and this relationship is statistically significant $p= .09$ only at the 0.1 level. In the second and the third study, the data analysis, did not reveal any relationships between personality traits and academic experience (see Table 2).

**Table 2.** Relation between personality traits and academic performance (1st study: N=19, 2nd study: N=80, 3rd study: N=46)

|  |  | 1st study | | 2nd study | | 3rd study | |
| --- | --- | --- | --- | --- | --- | --- | --- |
|  |  | Lab grade | Final grade | Lab grade | Final grade | Lab grade | Final grade |
| Extraversion | r | -.164 | -.039 | -.103 | -.027 | -.040 | -.100 |
|  | p | .503 | .873 | .361 | .812 | .743 | .411 |
| Agreeableness | r | -.262 | -.399 | .005 | .144 | -.003 | .086 |
|  | p | .279 | *.091* | .966 | .203 | .981 | .480 |
| Conscientiousness | r | -.021 | -.333 | .099 | .093 | .143 | .126 |
|  | p | .931 | .164 | .381 | .413 | .236 | .298 |
| Emotional Stability | r | -.009 | .025 | .161 | -.004 | -.147 | -.019 |
|  | p | .970 | .920 | .154 | .974 | .226 | .879 |
| Openness to experience | r | -.349 | -.004 | -.072 | .069 | -.014 | .026 |
|  | p | .143 | .987 | .527 | .544 | .909 | .833 |

**RQ3: relationship between Twitter activity and students' academic performance**

In the first study, each student tweeted 32.1 times on average (min=2, max=172, SD=37.30). This differentiation could be partially justified, by considering the optional participation to the procedure. The greatest activity was observed to the comments to other students' tweets (mean=21.42, SD=34.04), in comparison to the tweets made for the educational weekly activities (mean=8.9, SD=4.8) or tweets-questions related to the course (mean=1.7, SD=1.9). In the second study, each student tweeted 29.6 times on average (min=1, max=384, SD=44.5). The greatest interaction with the tool is noticed to the comments of fellow students' tweets (mean=18.6, SD=40.7) against the tweets made for the educational weekly activities (mean=9.8, SD=7.0) or tweets-questions related to the course (mean=1.0, SD=1.5). Regarding the tweets related to weekly activities the mean is lower since a specific number of tweets was requested. In the third study, each student tweeted 104.9 times on average. However, the data analysis showed a variation regarding the number of comments to fellow students with a max=1163 tweets and a min=2 tweets (SD=226.6). The greatest interaction with the tool is noticed to the comments of fellow students' tweets (mean=92.7, SD=221.2) against the tweets regarding the lab (mean=10.2, SD=7.3) and the questions related to the course (mean=1.8, SD=2.6).

**Table 3.** Relations between twitter activity and academic performance (1st study: N=19, 2nd study: N=80, 3rd study: N=46)

|  |  | 1st study | | 2nd study | | 3rd study | |
|---|---|---|---|---|---|---|---|
|  |  | Lab score | Final score | Lab score | Final score | Lab score | Final score |
| **Twitter Total** | r | .115 | .310 | .063 | .159 | -.119 | .132 |
|  | p | .639 | .196 | .579 | .159 | .432 | .381 |
| **Twitter Questions** | r | .039 | .490 | .166 | .049 | .055 | .229 |
|  | p | .873 | **.033** | .142 | .666 | .717 | .126 |
| **Twitter Activities** | r | .013 | .290 | .154 | .117 | -.008 | .119 |
|  | p | .957 | .228 | .174 | .301 | .956 | .430 |
| **Twitter Comments** | r | .122 | .271 | .043 | .150 | -.123 | .129 |
|  | p | .619 | .262 | .703 | .185 | .417 | .394 |

This result, observed in all studies, indicates the preference of students to interact between them rather than with their tutor, as proved by the frequencies of the different types of tweets (a. tweets for the weekly educational activities, b. tweets-questions set by the students to the tutor, c. tweets as comments to other tweets).

The 19 students who participated in the *first study* had a higher laboratory grade (mean=9.05/10, SD= .39) than the 17 students who did not participated (mean=8.12/10, SD=1.58). The same applies to the final grade of the course (mean=7.0/10, SD= 1.19) and (mean=5.5/10, SD=2.6) accordingly. Twitter activity refers to the students' activity regarding the number of tweets they posted related to the educational activities, their comments on other tweets, their questions set to the tutor and the total number of tweets. A relationship emerged between Twitter activity and the course's final grade ($r= .49$, $p= .03$).

The students who participated in the *second study* had a higher laboratory grade (mean=8.0/10, SD=1.50). than the 152 students who did not participated (mean=7.52/10. SD=1.74). The same applies to the final grade of the course (mean=5.1/10, SD=2.2 and mean=4.4/10, SD=5.1). However the statistical analysis did not show any significant relationship between Twitter activity and students' academic performance in the context of the course.

The students who participated in the *third study* had a higher laboratory grade (mean=8.8/10, SD=0.7) than the 24 students who did not participated (mean=8.3/10, SD=1.0). The same applies to the final grade of the course (mean=5.6/10, SD=2.1 and mean=5.3/10, SD=2.1). The data analysis did not unveil any significant relationships between Twitter activity and academic performance (Table 3).

**RQ4: Is there a difference between before and after the implementation of the educational activities via Twitter regarding: a) the attitude of the higher education students towards learning via social media b) the attitude of higher education students towards learning via ICT c) their self-reported attraction towards technology.**

The students' answers were recorded before and after their involvement with Twitter. The variables studied were: a) the attitude of the higher education students towards learning via social media by using Twitter in educational context. b) the attitude of higher education students towards learning via ICT and c) the self-reported attraction they feel towards technology.

In the first study, a t- test paired was conducted to compare the means in the students' attitudes prior conducting the activities and after their completion. Table 4 shows no differentiation regarding the attitude of "learning via social media" and of "self-reported attraction towards technology". A difference of 0.29 points exists in the attitude "ictl" (mean$_{pre}$= 3.60, mean$_{post}$=3.89, $p$= 0.045, s). In the second study, a t- test paired was conducted to compare the means. Table 4 shows a difference of 0.22 points in the attitude towards learning via social media (mean$_{pre}$= 3.28, mean$_{post}$=3.50, $p$= 0.006, s). In the third study, no statistically significant differences were observed.

**Table 4.** Difference in attitudes prior (Pre) and after (Post) conducting the study (1$^{st}$ study: N=10, 2$^{nd}$ study: N=73, 3$^{rd}$ study: N=46)

|  | 1$^{st}$ study | 2$^{nd}$ study | 3$^{rd}$ study |
|---|---|---|---|
| Learning via ICT Pre (mean) - Post (mean) $p$ | 3.60 – 3.89 **0.045** | 3.83 – 3.81 0.699 | 3.73 - 3.73 0.995 |
| Learning via social media Pre (mean) -Post (mean) $p$ | 3.23 – 3.23 1.000 | 3.31 – 3.51 **0.006** | 3.42 – 3.50 0.387 |
| Self-reported attraction towards Technology Pre (mean) - Post (mean) $p$ | 2.90 – 2.89 0.902 | 2.94-2.98 0.471 | 3.82 – 3.88 0.499 |

**Discussion and conclusion**

Twitter's integration in the educational process was examined in this study. Twitter has been used in the laboratory section of two academic courses for 8 weeks per each study. The students had to follow specific rules set by the researchers regarding the Twitter use.

The results suggest that the students who participated to every week's activities on Twitter, were more efficient in class compared to those who did not participate (both in the lab and in the final exams). However, no significant differentiations among the members of the first group was observed, according to their relative activity. Similar results were presented by Junco, Elavsky and Heiberger (2013), who showed that the inclusion of Twitter in the educational process led to an increase to the students' grade. On the other hand, they showed that the students that used Twitter in an optional base did not appear to have any increase to their grades.

Conscientiousness and openness to experience were related to students' Twitter activity in one study. This is partially in line with results reported by Chamorro-Premuzic & Furnham (2003a, 2003b) which underline the impact of conscientiousness on academic activity and performance. However, no relationship between the students' personality traits and their grade was unveiled contrary to the results of the beforementioned studies and those reported by Altanopoulou & Tselios (2015).

The greatest amount of activity in all studies was detected in the process of tweets' commenting. The number of tweets related to questions about the lab was small, but this number has been increased in the second study. This was a result of the students' preference for comments and interaction among them, rather than with their tutor. Considering the tweets regarding the weekly activities, the mean is smaller since the number of weekly tweets allowed was predetermined. Moreover, the results of the first and the second study showed a difference on the students' attitudes regarding "learning via ICT" and "learning via social media". This shows that the use of Twitter could change the students' attitudes regarding sharing information through social networks, searching information and including Web 2.0 in the academic community.

All in all, Twitter use helped the students to interact with their tutor, to communicate with their peers, enhanced their learning through the posts of their fellow students, motivated them to study the material of the class to answer weekly activities, to formulate and exchange ideas, and to familiarize with social networks. Therefore, the contribution of the present study to the existing research, is that proper Twitter mediated activities could enhance the learning experience. While the reported results could not be considered as generalizable and conclusive, it is argued that they constitute a necessary first step towards deeper understanding of the learning phenomena which take place in the context of Twitter mediated activities. In addition, by taking into consideration the personality characteristics of each student and how these characteristics can affect learning, the educator could better identify important aspects which promote collaboration in such a context.

Future work could build upon the findings of this study to investigate multiple directions. More specifically, studies with a broader and more diverse profile of participants are proposed. It is also proposed to conduct longitudinal studies to closer monitor students' possible variations in their attitudes towards social media adoption in education. In the present study, it is expected that there is no significant variation in students' perception for their ability to use the Internet. However, the use of the tool for a longer period is likely to significantly improve the students' self-efficacy (Hsu & Chiu, 2004).

Moreover, ways to control for the effect of potential impact of the grade incentives given to the students should be investigated. One obvious approach could be to lower or eliminate grade bonus. In addition, participation in Twitter activities will be a mandatory part of the course in subsequent academic years. Thus, differentiations between compulsory and voluntary participation will be also investigated. Finally, it would be interesting if a more systematic and deeper analysis of the type of students' interventions and cognitive strategies elicitation was conducted by using Educational Data mining and Learning Analytics techniques (Katsanos, Tselios & Avouris, 2008; Kotsiantis, Tselios, Filippidi & Komis, 2013; Tselios & Avouris, 2003).